\documentclass[]{pasj01}
\draft

\Received{}%{yyyy/mm/dd}
\Accepted{}%{yyyy/mm/dd}
 \usepackage{mathpazo}
 \usepackage[T1]{fontenc} 
\usepackage[utf8]{inputenc}
\begin{document} 

\title{ 
%\LETTERLABEL %%% <-- uncomment for LETTER article  
%\REVIEWLABEL %%% <-- uncomment for REVIEW article  
The reliability of the Titius-Bode relation and its implications for the search for exoplanets.}

%%% begin:list of authors
% Do NOT capitalize all letters in "textsc".
\author{Patricia \textsc{Lara}\altaffilmark{1}, Guadalupe \textsc{Cordero-Tercero}\altaffilmark{2}
\thanks{gcordero@igeofisica.unam.mx}}

\altaffiltext{1}{Posgrado en Ciencias de la Tierra, UNAM, Instituto de Geofísica-Circuito de la investigación Científica s/n, Ciudad Universitaria, 04150 Coyoacán, CDMX}
\email{lara@geofisica.unam.mx}

\altaffiltext{2}{Instituto de Geofísica, UNAM -Circuito de la investigación Científica s/n, Ciudad Universitaria, 04150 Coyoacán, CDMX}
\email{gcordero@igeofisica.unam.mx}

%\author{}
%\altaffiltext{2}{Instituto de Geofísica-Circuito de la investigación Científica s/n, Ciudad Universitaria, 04150 Coyoacán, CDMX}
%\email{gcordero@igeofisica.unam.mx}

\author{Christine \textsc{Allen}\altaffilmark{3}}
\altaffiltext{3}{Instituto de Astronomía, UNAM-Circuito de la investigación Científica s/n, Ciudad Universitaria, 04150 Coyoacán, CDMX}
\email{chris@astro.unam.mx}
%%% end:list of authors

%% `\KeyWords{}' always has to be placed before ``\maketitle'' 
%%  List of Key Words:  https://academic.oup.com/pasj/pages/Pasj_Keywords 

\maketitle

\begin{abstract}
The major semiaxes of the planets in our Solar System obey a simple geometric progression known as the Titius-Bode Relation (TBR), whose physical origin remains disputed. It has been shown that the exoplanetary systems follow a similar (but not identical) progression of the form $a_n= a_0$ e$^{bn}$, where $a_0$, $b$ are constants to be determined for each system. Since its formulation, the Titius-Bode Relation has proved to be highly predictive in our Solar System. Using data from 27 exoplanetary systems with 5 or more planets and applying a proposed method, we conclude that reliable TB-like fits can be obtained for systems with at least 4 planets and that the precision of the TBR is $78\%$. By means of a statistical test we show that the periods of planets in real exoplanetary systems are not consistent with a random distribution. Rather, they show signs that their configuration is shaped by their mutual interactions.
\end{abstract}

\section{Introduction}
Although the Titius-Bode law has now been known for over 200 years (see \cite{Nieto1972}, for a historical account), its physical basis remains largely unclear. While some argue that it is entirely coincidental, others attribute it to the physical conditions in the initial proto-planetary disk \citep{Graner1994} \citep{Hayes1998} \citep{Isaacman1977} or to a dynamical relaxation process \citep{Dermott1968} \citep{Dermott1969} \citep{Hills1970} \citep{Neslusan2004} \citep{Ovenden1975}. The TB \textquotedblleft law \textquotedblright is not really a law in the physical sense, but rather a numerical relation. However, since the term \textquotedblleft Titius-Bode law\textquotedblright  has been widely adopted throughout the years, we will use \textquotedblleft TB law\textquotedblright  and \textquotedblleft TB relation\textquotedblright  interchangeably.\\

\noindent In its classical form, the Titius-Bode (TB) law states that the major semiaxes $a$ (measured in AU) of the planets in our Solar System obey a simple geometric progression of the form:

\begin{equation}
  a_n=0.4+0.3\times 2^n,
\end{equation}

\noindent where the index $n$ is $- \infty $ for Mercury, 0 for Venus, 1 for the Earth, etc. Interestingly, the major satellites of Jupiter, Saturn, and Uranus obey a similar (but not identical) progression, known as Dermott's law (\cite{Dermott1968}, \cite{Dermott1969}). Dermott's law is a relation between the orbital periods of the satellites and takes the form:

\begin{equation}
  p_n=p_0\times k^n,
\end{equation}

\noindent where $p_0$ and $k$ are constants to be determined for each system, and $n$ is 0 for the innermost satellite, 1 for the next satellite, etc. For Keplerian orbits, a system obeying Dermott's law will obey a similar law for the major semiaxes:

\begin{equation} \label{ec3}
    a_n=a_0 \times C^n,
    \end{equation}

\begin{equation} \label{ec4}
   a_n = a_0 e^{bn},
\end{equation}
\noindent where $a_0$ is the major semiaxis of the planet nearest to the star and $n=0, 1, ...$.\\

\noindent As shown by \citet{Poveda2008}, the period distribution of the planets in the Solar System and 55 Cancri can be described by equation \ref{ec4}. According to these authors, the quality of the fit provided by equation \ref{ec4} is poorer than that provided by the TB relation (Eq. 1) for the inner seven planets, but better for Neptune. The fact that the planets in the Solar System and the exoplanetary systems follow a similar structural law does suggest that such laws are more than  mere coincidences.\\

\noindent \citet{Lara2012} and \citet{Poveda2008} have shown that exoplanetary systems, like the Solar System itself, follow a Titius-Bode relation. Historically the relation was used as a guide for discovering planets in the Solar System \citep{Nieto1972}; hence we propose that it could be a good guide to discover additional exoplanets.\\

\noindent \citet{Bovaird2013}, using a sample of 68 exoplanetary systems with at least 4 planets (both confirmed and candidates),  showed that the TB relation can be useful for making predictions about the periods and positions of exoplanets. To do this, they inserted some vacancies and found that the best fit was the one with the highest $\gamma$, where $\gamma$ is a measure of the improvement in the $\chi^2$/(degrees of freedom) per inserted planet. In addition, to assess the completeness of their systems (including the inserted vacancies) they used a dynamical spacing $\Delta$ (an estimate of the stability of adjacent planets). They concluded that the TB relation can be useful to find new planets in exoplanetary systems.\\

\noindent The purpose of this article is to test the predictive capacity of the TB relation using a kind of “trick”. Our sample contains 27 confirmed exoplanetary systems with at least 5 members. To test the TB relation we perform fits with the first 4 confirmed planets and test how well our method reproduces the periods of the other known members of the system. This idea is based on the historical discovery of Ceres, Uranus and Neptune, since the classical TB relation was able to predict them. In other words, we will test the TB relation using known planets; this approach is different from that of \citet{Bovaird2013}. \\

\section{Method}

Our hypothesis is that if we have at least 4 observed orbital periods of planets in an exoplanetary system, then the TB relation will be able to predict other planets. To test this hypothesis we use a sample containing all known extrasolar systems with 5 or more planets. For each system, we use the first four discovered planets to obtain a TB relation and to predict the periods of other possible planets. The results are compared with the real planet(s) discovered later. We do so in order to obtain a reliable assessment of the capacity of the TB relation to find new planets in exoplanetary systems. For extrasolar systems in which the planetary discoveries were reported on different dates, the chronology is clear, but for systems reported at the same time, we use the four smallest periods.\\

\noindent We had to decide the minimum number of planets that could be used to apply the TBR to predict new objects. We found that if we took 3 planets, the errors were very large (28-$140\%$). The errors decreased appreciably using at least 4 planets (the error is calculated by $ ((p_{pred} -p_{obs})/p_{obs})\times 100\%$, where $p_{pred}$ is the period calculated by the TB fit and $p_{obs}$ is the observable data). This is the reason for using at least 4 planets to find a reliable relation. For systems with more than 5 planets we applied the method in the way we describe in the next paragraphs.\\

\noindent In this analysis we will use orbital periods instead major semiaxes to avoid the error introduced by the uncertain distance and mass of the primary star, so that\\

\begin{equation}
\label{exp}
  p_n=p_0e^{(bn)},
\end{equation}

\noindent where $n$ is the number in ascending order of the planets in a planetary system, $p_0$ and $b$ are constants to be determined from the fits, and $p_n$ are the observed periods.\\

\begin{figure}
  \begin{center}
    \includegraphics[scale=0.65]{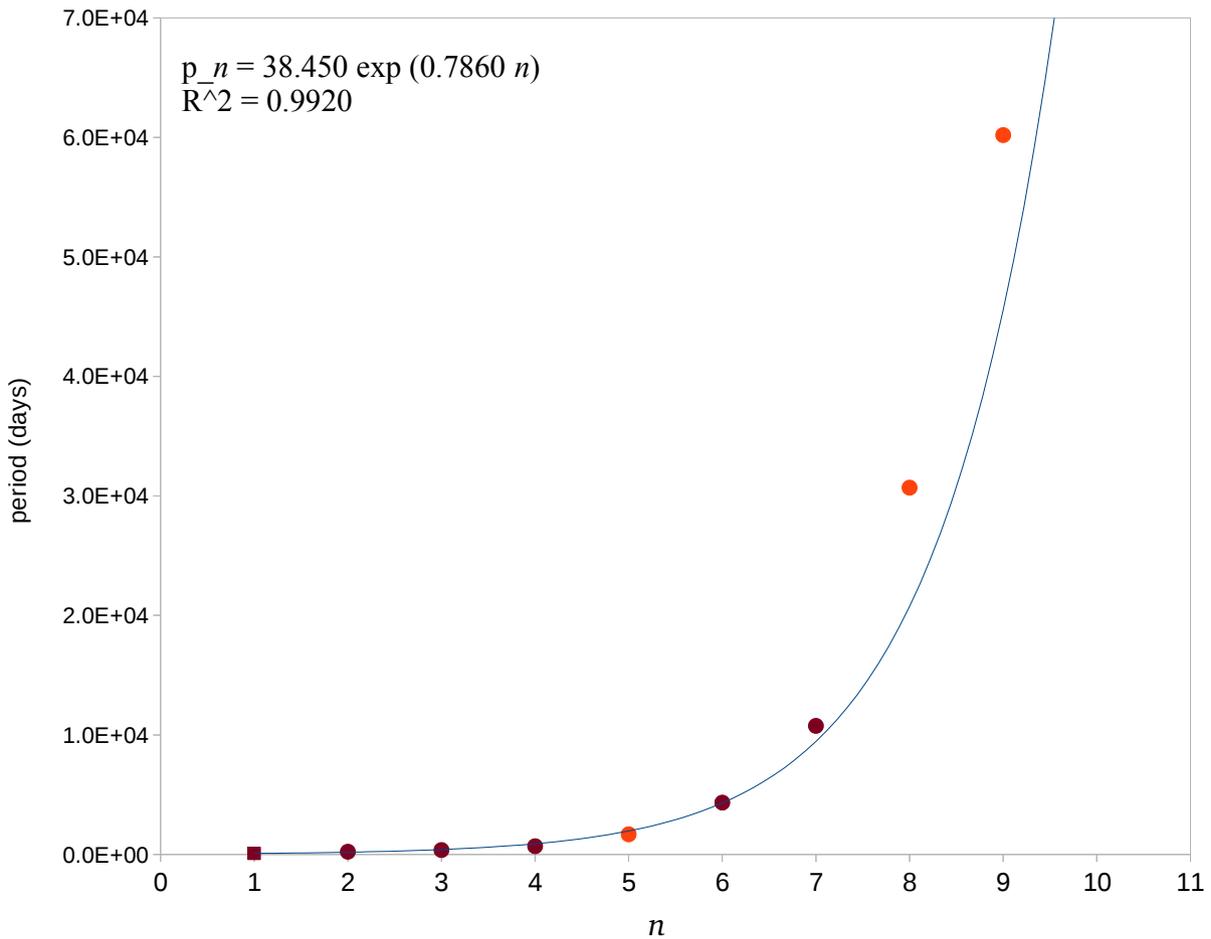}
        \caption{Fit to the planets of the Solar System that were discovered before Ceres, Uranus and Neptune.}\label{fig:1}
       \end{center}
\end{figure}

\noindent To apply our method we need a fit that can be used as a reference. We used the fit to the planets of the Solar System done by Titius, who postulated that there was an unoccupied space (a vacancy) between Mars and Jupiter.  This way of treating the data proved to be a good method to predict new planetary bodies. So first, we made a fit similar to the Titius law, that is, we fitted equation 5 to Mercury, Venus, Earth, Mars, Jupiter, Saturn and the vacancy between Mars and Jupiter. The coefficient of correlation we obtain is 0.9920. The relationship found predicted the periods of Ceres, Uranus and Neptune with errors of $17\%$, $32\%$ and $24\%$, respectively (Figure \ref{fig:1}). We assumed then that in the case of exoplanets the vacancies found could represent places where planets, asteroids or belts of minor planets may exist.\\

%\selectlanguage{english} 

\begin{figure}
  \begin{center}
    \includegraphics[scale=0.5]{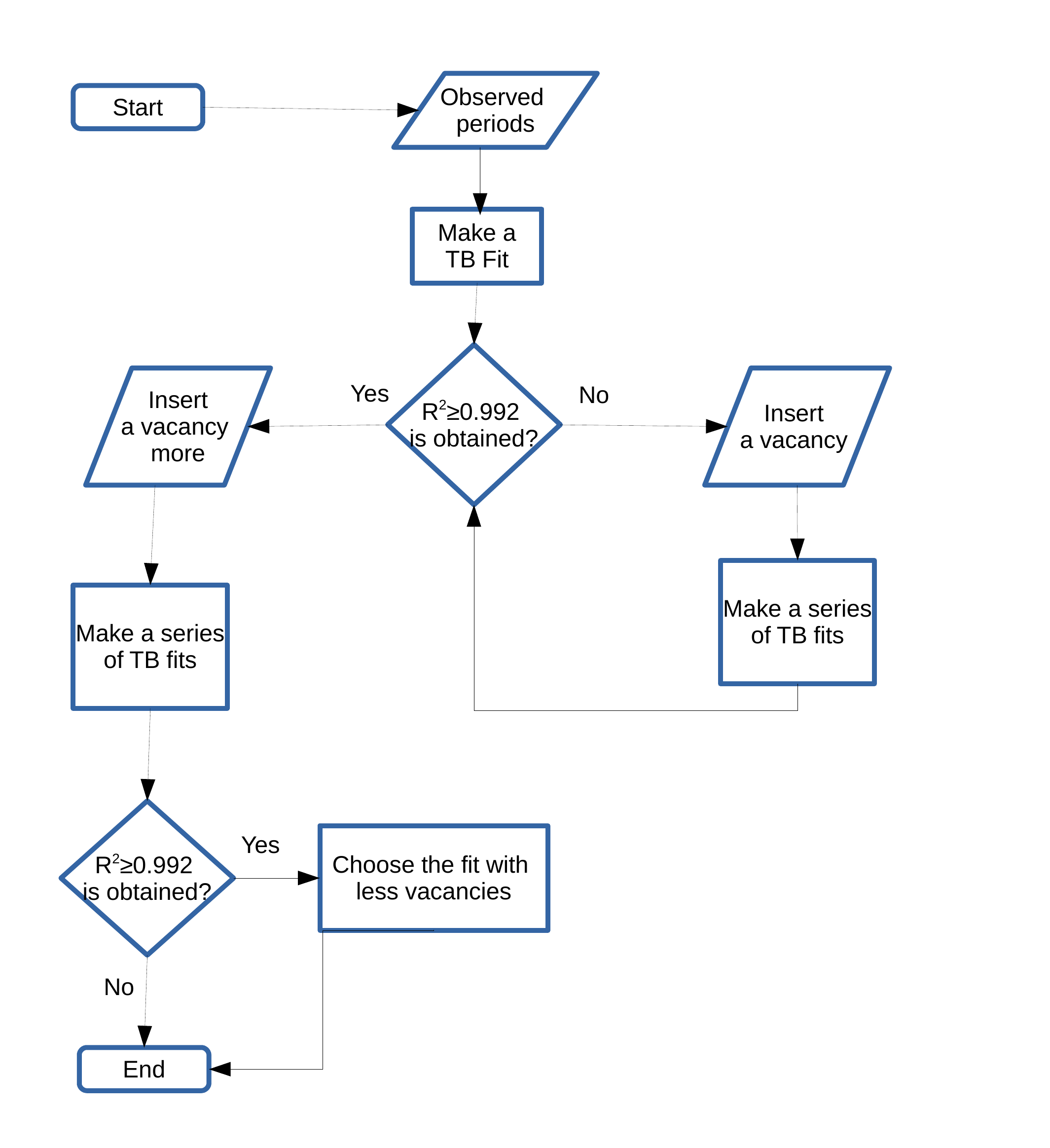}
        \caption{Flow chart of the applied method.}\label{fig:2}
       \end{center}
\end{figure}

\noindent To show that this method is a good approach to predict the existence of possible planets in other planetary systems, we used the 27 known systems with 5 or more exoplanets and proceeded as follows: we considered the first four discovered planets of each of the systems in our sample and we fit an exponential TB relation (Equation 5) assuming no vacancies. If that fit did not result in a correlation coefficient of  R$^2 = 0.9920$ or greater, a series of fits were made introducing vacancies between planets and considering that the first observed planet always has $n=1$. To do this, we tested all the combinations with one vacancy between planets; if one of these combinations with one vacancy fulfilled the criterion $R^2 \geq  0.9920$, we tested all the possible combinations with two vacancies to check if there were other combinations that fulfilled the criterion. In all cases we found no fit with two vacancies that fulfilled the criterion $R^2\geq 0.9920$. In the cases where the criterion $R^2$ was not fulfilled with a one vacancy but with two vacancies, we tested all possible combinations with three vacancies to check whether another, better combination could be found. As in the previous case, the only fit that fulfilled the criterion was the one with two vacancies. This process was repeated when a better fit was obtained with three and four vacancies. It is important to note that in all the cases there was one and only one combination that fulfilled the criterion and it was the one with the smallest number of  vacancies. In Figure \ref{fig:2} we show a flow graph our method. The number of possible combinations when vacancies are inserted is given by equation \ref{comb}.\\

\begin{equation}
  n=\frac{(Vacancies +(Observed-2)!}{(Observed-2)!(Vacancies)!}.
\label{comb}
\end{equation}

\noindent We considered our method to be successful if:
\begin{enumerate}
  \item The same vacancies remain over the iterations, and
  \item The fit predicts the next discovered planet (s) with an error less than $32\%$.
\end{enumerate}

\noindent The first condition gives us a confidence about the method because if the vacancies did not persist the method would not be useful. The second condition is mandatory to show that the method is actually useful to find real planets.

\section{Probability of success of the Titius-Bode Relation}

\noindent Our method was applied to the sample of the 27 available exoplanetary systems that house at least 5 planets. All the bodies in these systems are confirmed. To corroborate the data we used the 3 exoplanet databases (https://exoplanetarchive.ipac.caltech.edu, http://exoplanet.eu and http://openexoplanetcatalogue.com). The 27 confirmed exoplanetary systems with 5 planets or more are: 55 Cancri, HD 40307, HIP 41278, K2-138, Kepler-102, Kepler-122, Kepler-150, Kepler-154, Kepler-169, Kepler-186, Kepler-238, Kepler-292, Kepler-296, Kepler-33, Kepler -444, Kepler-55, Kepler-62, Kepler-84, GJ 667C, HD 10180, HD 219134, HD 34445, Kepler-11, Kepler-20, Kepler-80, Trappist-1 and Kepler-90. \\

\noindent After analyzing the 27 exoplanetary systems, we found that the proposed method gave successful results for 21 systems, i.e. that the precision of our method is 78\% (using the standard definition of precision as the ratio between the number of successful cases and the total number of cases). The proposed method was applied successfully to the exoplanetary systems 55 Cancri, HD 40307, HIP 41378, K2-138, Kepler-102, Kepler-154, Kepler-186, Kepler-238, Kepler-296, Kepler-33, Kepler-444, Kepler-62, Kepler-84, GJ 667C, HD 34445, Kepler-20, Kepler-11, Kepler-80, HD 219134, Trappist-1 and Kepler-90. In these planetary systems, the vacancies we found coincided with planets that were later detected. In the next paragraphs we describe in detail 7 cases, as examples. \\

\noindent Kepler-102 was one of the exoplanetary systems where the predictive capacity of the TBR was very clear. In this case the vacancy generated by our method coincided with the orbital period of the later found planet (Kepler-102 f) with an error of $10.43\%$. In addition, when we applied the method to all observed planets no further vacancies were found.\\

\noindent For Kepler-20, the fit for the first four planets showed 2 vacancies that would correspond to planets with periods of 21.75 and 40.29 days. These periods coincide with the periods of the planets Kepler-20f and Kepler-20g with errors of $11.08\%$ and $15.31\%$, respectively. When the fit was made with 5 observed planets, we obtained a vacancy that coincided with the period of the later found planet (Kepler-20g) with an error of $12\%$. \\

\begin{figure}[h!]

  \begin{center}
    \includegraphics[scale=0.45]{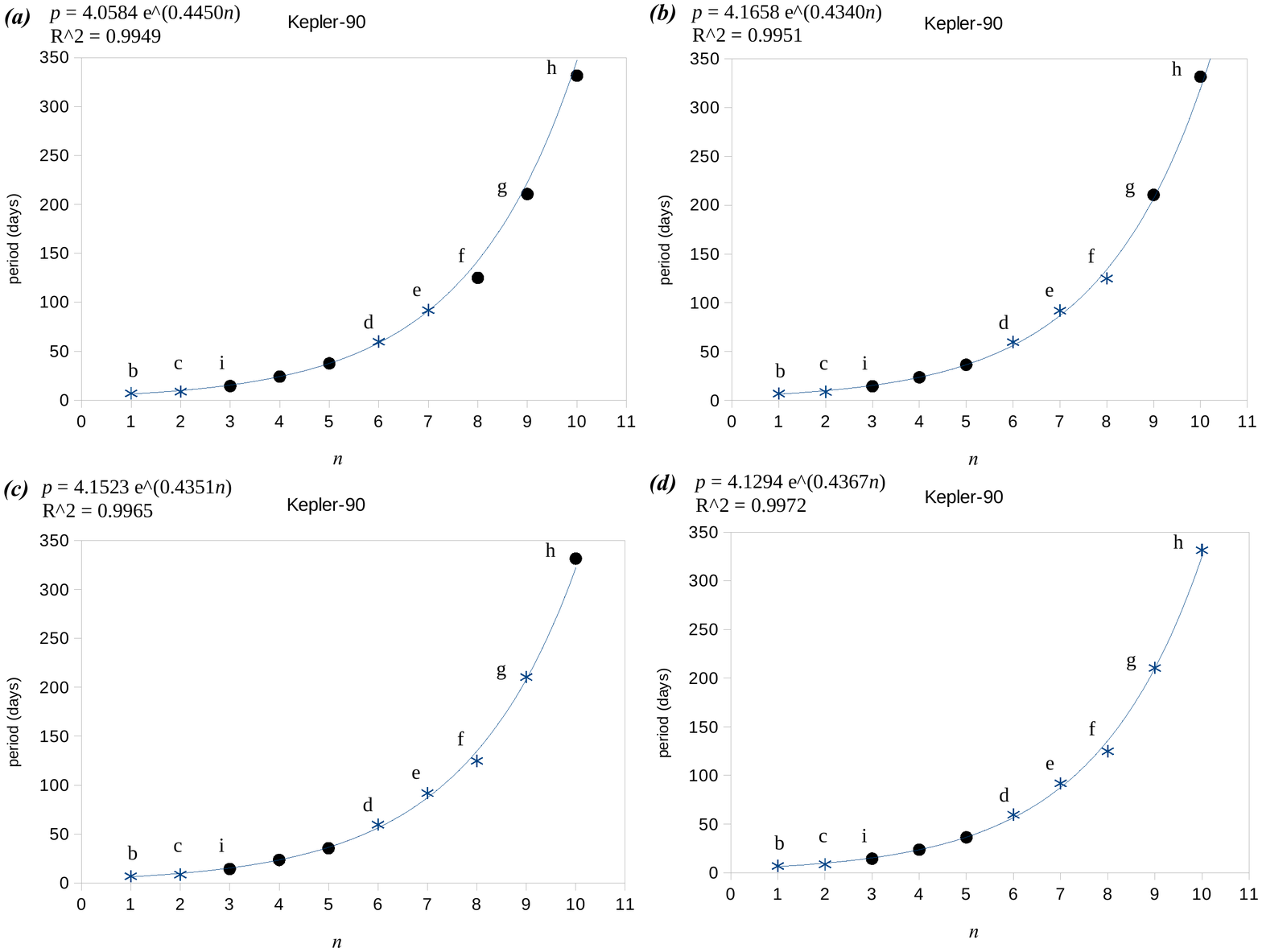}\\
    \includegraphics[scale=0.4]{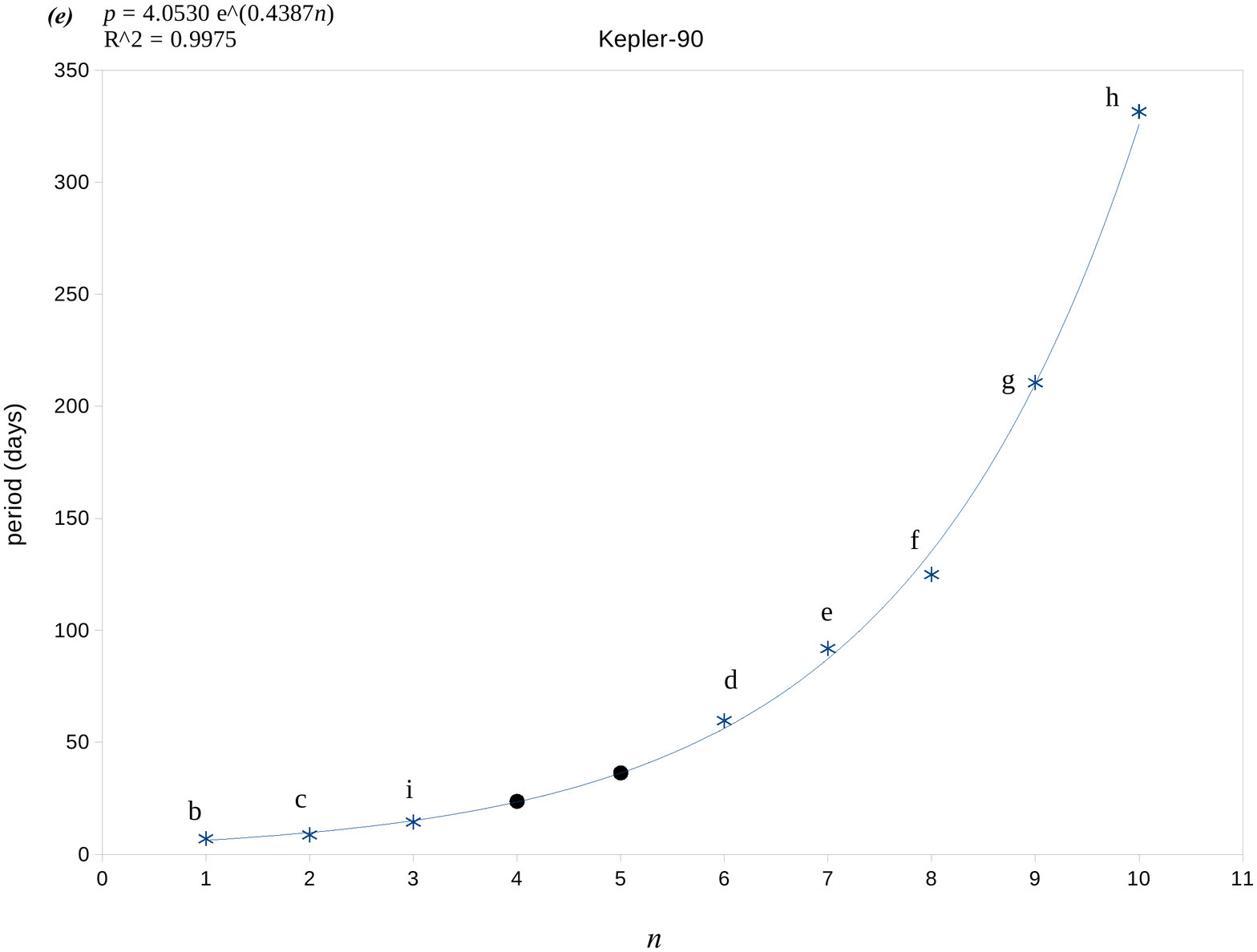}
        \caption{Method applied to the planetary system Kepler-90. Asterisks denote the planets used for the fit, and circles the vacancies. (a) Fit using the first four known planets. Note the vacancies at $n=3$, 4 and 5, and the extrapolation to $n=8$, 9 and 10, which coincides with the most external planets. (b) Fit using the first five known planets. Note that the vacancies at $n=3$, 4 and 5 remain as well as the extrapolations. (c) Fit using the first six known planets. Vacancies and extrapolations remain. (d) and (e) Fits using 7 and 8 planets. As in the other cases, the vacancies are conserved. In addition, the vacancy at $n=3$ coincides with Kepler-90i.}
         \label{fig2}
       \end{center}
       
 \end{figure}

\noindent Before December 2017, Kepler-90 had 7 known planets. The fit with the first four planets gave three vacancies. Even though none of them matched the other three observed planets, when the TB relation was extrapolated we found the periods of another three planets, later discovered (Kepler-90f, g and h), with errors of $14.3\%$, $5.8\%$ and $4.3\%$, respectively. If we use 5 planets for the fit, the previous three vacancies remain, and the next 2 extrapolated planets coincide with the orbital periods of Kepler-90g and h with errors of $1.7\%$ and $3.6\%$. Repeating the method using six planets for the fit, the vacancies remain and the extrapolated planet coincides with Kepler-90h with an error of $3.6\%$. The TB relation indicated that they were 3 putative objects to be discovered in this planetary system. In december 2017, the discovery of Kepler-90i was reported (\citet{Shaulle2018}). It coincides with one of the vacancies that we found, with an error of 6.0 \%. When we use these eight observed planets the other  two vacancies remain. This   supports the existence of two possible planetary bodies in the two vacancies (Figure \ref{fig2}).\\

\noindent The exoplanetary system Trappist-1 contains 7 planets. Using planets b, c, d and e, for the fit, the result did not show vacancies, but when extrapolated, the relation showed 3 further orbitals, which coincide with Trappist-1 f, g and h with errors of $8.42\%$, $29.3\%$ and $27.8\%$, respectively. When the first 5 planets were taken for the fit, no vacancies appeared, but when extrapolated, 2 orbital spaces were obtained coinciding with the g and h planets with errors of $21.20\%$ and $17.85\%$, respectively. The fit using 6 observed planets predicted an orbital period of 20.73 days that coincided with the orbital period of Trappist-1 h with an error of $3.65\%$.\\

\noindent Other examples where the TB predictive capacity was remarkable were Kepler-296, Kepler-444 and K2-138. They did not  show vacancies but, when extrapolated, we obtained in all three cases the fifth planet with errors of $0.63\%$, $4.23\%$ and $1.86\%$, respectively. These examples show that vacancies and extrapolations of the TB relation are good indicators of the probable existence of planets.\\

\noindent The fits to  55 Cancri, HD 40307, HIP 41378, Kepler -154, Kepler- 186, Kepler-238, Kepler-33, Kepler-62, Kepler-84,  GJ 667C, HD 34445, Kepler-11, Kepler-80 and HD 219134 showed vacancies that may indicate the orbital periods of undiscovered planets or asteroids belts. We found that the same vacancies persisted in all cases, which strengthens our conclusion.\\

\noindent All the previous examples show that the TB relation can be used as a guide for further detections. The predicted periods and their errors are displayed in Table \ref{Table1}. The variables $p_0$, $b$, R$^2$ are defined as in equation 5. On average, a percentage error of $9.84\%$ between the predictions and the observed planets was found for the 21 systems studied (Table \ref{Table1}). It is interesting that when fits were made with all the observed planets, the number of vacancies persisted, so we expect that they indicate a possible future discovery of a planet, a minor planet or an asteroid belt.\\

\noindent The remaining 6 systems were unsuccessful cases because the vacancies did not persist, that is, when we used different numbers of planets for the fit, different vacancies showed up. In addition, the systems Kepler-122 and HD 10180 showed errors greater than $30\%$. The predictions for these cases are shown in Table \ref{Table2}.\\

\begin{footnotesize}
  
\begin{center}
\begin{longtable}{lcccccccccc}

\caption{Predicted periods and errors for the systems with the best fits.}\label{Table1}\\
\hline
\footnotesize Planetary&\footnotesize Number&\footnotesize Number&\footnotesize Extrapolatios&$p_0$&$b$&$R^2$&$\chi^2$&\footnotesize Predicted& \footnotesize Observed& \footnotesize error \\
system & of planets & of vacancies & &  &&&&periods&period& $\%$ \\
 & & & & & & & & (days)&(days)& \\ 
\hline
\endfirsthead
\multicolumn{11}{c}%
{\tablename\ \thetable\ -- \textit{Continued from previous page}} \\
\hline
\footnotesize Planetary&\footnotesize Number&\footnotesize Number&\footnotesize Extrapolations&$p_0$&$b$&$R^2$&$\chi^2$&\footnotesize Predicted&\footnotesize Observed&\footnotesize error \\
system & of planets & of vacancies & &  & & & &periods&period& \% \\
 & & & & & & & &(days)&(days) & \\ \hline
\endhead
\hline \multicolumn{11}{c}{\textit{Continued on next page}} \\
\endfoot
\hline
\endlastfoot
55 Cnc & 4 & 3 & 0	& 0.1594 & 1.4717 & 0.9978 & 48.28 &	3.0	&		&		\\
&		&		&		&&&&&	250.3	&	260.7	&	3.98	\\
&		&		&		&&&&&  1090.6	&		&		\\
 & & & & & & & & & & \\
HD 40307	&	4	&	0	&	2	& 1.8589 & 0.8209 & 0.9981 & 0.1857 &	112.7	&		&		\\
	&		&		&&&&&		&	256.0	&	198.8	&	23.79	\\
 & & & & & & & & & & \\
HIP 41378	&	4	&	3	&	2	& 10.1334 & 0.4041 & 0.9920 & 3.7813& 22.7	&		&		\\
	&		&		&&&&&		&	51.0	&		&		\\
	&		&		&&&&&		&	76.4	&		&		\\
		&		&		&&&&&		&256.98		&		&		\\
	&		&		&&&&&		&	384.8	&	324	&	18.77	\\
 & & & & & & & & & & \\	
K2-138 & 4 & 0 & 1	& 1.5447 & 0.4158 & 0.99997 & 0.00088 &	12.5	&	12.7576	& 1.86		\\
 & & & & & & & & & & \\
Kepler-102	&	4	&	3	&	0	&4.0562 & 0.2793 & 0.9998 & 0.0041 & 9.2 &	10.31	&	10.43	\\
&		&		&		&&&&&	12.2	&	&	\\
&		&		&		&&&&&	21.0	&	&	\\	
 & & & & & & & & & & \\			
Kepler-154	&	4	&	2	& 0 	&	4.0562 	& 0.2793 	& 0.9998& 0.1726 &	6.8	&		&		\\
		&		&	&		&&&&&	11.7	&	9.92	&	17.85\\
 & & & & & & & & & & \\	
Kepler-186	&	4	&	0	& 3	& 2.2135 	&	0.5863	&0.9982	& 0.0435 & 	41.5	&		&		\\
			&	&		&&&&&	&	74.6	&		&		\\
			&	&		&&&&&	&	134.1	&	129.95	&	3.21	\\
 & & & & & & & & & & \\	
 Kepler-238	&	4	&	1	&	1	& 1.0884 &0.6125 & 0.9954 & 0.108 &	3.7	&		&		\\
	 		&	&			&&&&&		&	42.9	&	50.447	&	15.0	\\
			 & & & & & & & & & & \\	
 Kepler-296	&	4	&	0	&	1	& 3.2953 & 0.5899	& 0.9990	& 0.0329	&	62.9	&	63.3363	&	0.63	\\
 & & & & & & & & & & \\	
 Kepler-33	&	4	&	1	&	1	& 3.6544 & 0.4361 & 0.9984 & 0.0542 &	8.7	&		&		\\
				&&		&&&&&		&	50.0	&	41.029	&	22.00	\\
 & & & & & & & & & & \\	
 Kepler-444	&	4	&	0	&	1	& 2.7584 & 0.2606 & 0.9962 & 0.0074 &	10.2	&	9.7405	&	4.23	\\
 & & & & & & & & & & \\
 Kepler-62	&	4	&	2	&	1	& 3.2858 & 0.6010 & 0.9945 & 0.3959 &	36.4	&		&		\\
		&		&		&&&&&		&	66.3	&		&		\\
		&		&		&&&&&		&	220.7	&	267.291	&	17.43	\\
 & & & & & & & & & & \\
 Kepler-84	&	4	&	2	&	1	& 2.8778 & 0.3748 & 0.998 & 0.0033 & 	6.1	&		&		\\
		&		&		&&&&&		&	18.8	&		&		\\
		&		&		&&&&&		&	39.7	&	44.55	&	11.00	\\
 & & & & & & & & & & \\		
GJ 667C	&	4	&	3	&	2	& 4.8559 & 0.4244 & 0.9985 & 0.2097 &	11.4	&		&		\\
		&		&		&&&&&		&	17.3	&		&		\\
		&		&		&&&&&		&	40.5	&	39.026	&	3.87	\\
		&		&		&&&&&		&	144.8	&		&		\\
		&		&		&&&&&		&	221.4	&	256.2	&	13.60	\\	
 & & & & & & & & & & \\																
 	&		5	&	2	&	2 & 4.8351 & 0.4237 & 0.9982 & 0.2374	&	11.3	&		&		\\
	&			&		&&&&&		&	17.2	&		&		\\
	&			&		&&&&&		&	143.4	&		&		\\
	&			&		&&&&&		&	219.0	&	256.2	&	14.53	\\
	 & & & & & & & & & & \\
HD 34445&	4	&	2	&	4	& 26.9052 & 0.5273 & 0.9941 & 4.3382 & 77.2 &		&	\\
 &		&		&		&&&&&	375.7	&	&	\\
&		&		&		&&&&&	1078.6	& 1056.7	& 2.08\\
&		&		&		&&&&&	1827.6	&	&	\\
&		&		&		&&&&&	3096.6	&	&	\\
&		&		&		&&&&&	5246.7	& 5700.0	&	7.95\\	
 & & & & & & & & & & \\
&	5	&	2	&	3	& 27.0806 & 0.5248 & 0.9966 & 5.0904 & 77.4&	&	\\
&		&		&		&&&&&	373.5	&	&	\\
&		&		&		&&&&&	1803.0	&	&	\\
&		&		&		&&&&&	3047.3	&	&	\\
&		&		&		&&&&&	5150.3	& 5700.0	&	9.64\\	
 & & & & & & & & & & \\
Kepler-20& 4 & 2 & 0 & 1.8461 & 0.6166 & 0.9972 & 0.2161 & 21.8 &19.577&11.08\\
			&		&		&       &        &        &        &		&	40.3&	34.94	&	15.31	\\
			&		&		&       &        &        &        &		&		    &	     	&		\\
			&	5	&	1	&	0	& 1.8408 & 0.6109 & 0.9958& 0.6409 &	39.1	&	34.94	&	11.77 \\
 & & & & & & & & & & \\			
Kepler-11	&	4	&	3	&	0	& 6.2186 & 0.4138 & 0.9945 & 0.6136 &	21.5	&	22.6845	&	5.15	\\
			&		&		&&&&&		&	32.6	&	31.9996	&	1.70	\\
			&		&		&&&&&		&	74.4	&		&		\\
 & & & & & & & & & & \\								
 &	5	&	2	&		& 6.3157 & 0.4124 & 0.9941 & 0.640 &	32.9	&	31.9996	&	2.73	\\
			&		&		&&&&&		&	59.4	&		&		\\											
 & & & & & & & & & & \\	
 Kepler-80	&	4	&	3	&	1	& 0.6817 & 0.3823 & 0.9989 & 0.0278 &	1.5	&	&		\\
			&		&		&&&&&		&	2.2	&		&		\\
			&		&		&&&&&		&	3.2	&	3.07	&	2.41	\\
			&		&		&&&&&		&	14.5&	14.6456	&	0.89	\\
	 & & & & & & & & & & \\													&	5	&	2	&	1	& 0.6762 & 0.3830 & 0.9987 & 0.0294 &	1.5	&	&		\\
			&		&		&&&&&		&	2.1	&		&		\\											
			&		&		&&&&&		&	14.5	&	14.6456	&	1.15	\\											
 & & & & & & & & & & \\								
 HD 219134	&	4	&	4	&	0	& 1.0675 & 0.9404 & 0.9969 & 43.221 & 17.9	&	22.717	&	21.13	\\
	&		&		&&&&&		&	45.9	&	46.859	&	2.10	\\
	&		&		&&&&&		&	300.7	&		&		\\
	&		&		&&&&&		&	769.9	&		&		\\
 & & & & & & & & & & \\
	&	5	&	3	&		& 1.0721 & 0.9401 & 0.9969 & 40.9815 &	18.0	&	22.717	&	20.80	\\
	&		&		&&&&&		&	302.0&		&		\\
	&		&		&&&&&		&	773.2	&		&		\\
 & & & & & & & & & & \\
 Trappist-1	&	4	&	0	&	3	& 0.9520 & 0.4700 & 0.9981 & 0.0091 &	10.0	&	9.2067	&	8.42	\\
	&				&		&&&&&		&	16.0	&	12.3529	&	29.49\\
	&				&		&&&&&		&	25.5	&	20	&	27.77\\	
 & & & & & & & & & & \\											
	&			5	&	0	&	2	& 0.9835 & 0.4538 & 0.9977 & 0.0230 &	14.97	&	12.35	&	21.20	\\
	&				&		&&&&&		&	23.57	&	20	&	17.85	\\
 & & & & & & & & & & \\
  &	6	&	0	&	1	& 1.0486 & 0.4263 & 0.9930 & 0.1649 &	20.4	&	20	&	1.93	\\
 & & & & & & & & & & \\	
Kepler-90&	4&3	&	3& 4.0584 & 0.4450 & 0.9949 & 0.2334 &15.4	&14.4491	&	6.74\\
			&		&		&&&&&		&	24.1	&		&		\\
			&		&		&&&&&		&	37.6	&		&		\\
			&		&		&&&&&		&	142.8	&	124.91	&	14.30 \\
			&		&		&&&&&		&	222.8	&	210.61	&	5.77 \\
			&		&		&&&&&		&	347.5	&	331.6	&	4.83	\\
 & & & & & & & & & & \\												
			&	5	&	3	&	2	& 4.1658 & 0.4340 & 0.9951 & 1.332 &	15.3	&	14.4492	&	6.0\\
			&		&		&&&&&		&	23.6	&		&		\\
			&		&		&&&&&		&	36.5	&		&		\\
			&		&		&&&&&		&	207.0	&	210.61	&	1.66\\
			&		&		&&&&&		&	319.5	&	331.6	&	3.60\\
  & & & & & & & & & & \\															
Kepler-90			&	6	&	3	&	1	& 4.1523 & 0.4351 & 0.9965 & 1.7094 &	15.3	&	14.4492	&	6.0\\
			&		&		&&&&&		&	23.7	&		&		\\
			&		&		&&&&&		&	36.6	&		&		\\
			&		&		&&&&&		&	322.1	&	331.6	&	2.85	\\
  & & & & & & & & & & \\
			&	7	&	2	&	0	& 4.1294 & 0.4367 & 0.9972 & 1.5493 &	15.3	&	14.4492	&	6.0 	\\
			&		&		&&&&&		&	23.7	&		&		\\
			&		&		&&&&&		&	36.4	&		&		\\
\end{longtable}
\end{center}

\begin{center}
\begin{longtable}{lcccccccccc}
 \caption{ Predicted periods and errors for the systems with non-conserved vacancies}\label{Table2}\\
\hline
Planetary&Number&Number&Extrapolations&$p_0$&b&$R^2$&$\chi^2$&Predicted&Observed&error \\
system & of planets & of vacancies & &  &&&&periods&period& \% \\
 &  & & && &&& (days)&(days)& \\ 
\hline
\endfirsthead
\multicolumn{11}{c}%
{\tablename\ \thetable\ -- \textit{Continued from previous page}} \\
\hline
Planetary&Number&Number&Extrapolations&$p_0$&b&$R^2$&$\chi^2$&Predicted&Observed&error \\
system & of planets & of vacancies & &  &&&&periods&period& \% \\
& &  & & && && (days)&(days)& \\ \hline
\hline
\endhead
\hline \multicolumn{11}{c}{\textit{Continued on next page}} \\
\endfoot
\hline
\endlastfoot		
Kepler-122	&	4	&	0	&	1	& 3.3029 & 0.6205 & 0.9930 & 0.1825 &	73.5	&	56.268	&	30.6 	\\
  & & & & & & & & & & \\
Kepler-150	&	4	&	2 &	 7&  2.1119 & 0.4428 & 0.9967 & 0.06915 &    5.1  &	&	\\
	 &		&		&&&&&		&	19.3&		&		\\
	 &		&		&&&&&		&	46.9	&		&		\\
	 &		&		&&&&&		&	73.0	&		&		\\
	 &		&		&&&&&		&	113.6&		&		\\
	 &		&		&&&&&		&	177.0	&		&		\\
	 &		&		&&&&&		&	275.5	&		&		\\
	 &		&		&&&&&		&	429.0	&		&		\\
	 &		&		&&&&&		&	668.0	&	636.209	&	4.840 \\
	   & & & & & & & & & & \\
Kepler-169	&	4	&	2	&	6	& 2.5220 & 0.2890 & 0.9951 & 0.04317 &	4.5	&		&		\\
	&		&		&&&&&		&	10.7	&		&		\\
	&		&		&&&&&		&	19.1	&		&		\\
	&		&		&&&&&		&	25.5	&		&		\\
	&		&		&&&&&		&	34.0	&		&		\\
	&		&		&&&&&		&	45.4	&		&		\\
	&		&		&&&&&		&	60.6	&		&		\\
	&		&		&&&&&		&	80.9	&	87.09	&	7.06	\\
   & & & & & & & & & & \\													
Kepler-292	&	4	&	2	&	1	& 1.9707 & 0.3065 & 0.9961 & 0.0364 &	4.9	&		&		\\
	&		&		&&&&&		&	9.1	&		&		\\
	&		&		&&&&&		&	16.8	&	19.8066	&	15.00	\\
& & & & & & & & & & \\													
 Kepler-55	&	4	&	2	&	0	& 1.3171 & 0.5924 & 0.9955 & 0.6069 &	7.8	&	10.19	&	23.63	\\
	&		&		&&&&&		&	14.9	&		&		\\
& & & & & & & & & & \\
HD 10180	&	4	&	0	&	3	& 2.1026 & 1.0290 & 0.9983 & 0.5896 &	360.7	&	604.67	&	40.34	\\
	&		&		&&&&&		&	1009.4	&		&		\\
	&		&		&&&&&		&	2824.6	&	2205	&	28.10	\\
	&		&		&&&&&		&		&		&		\\
	& & & & & & & & & & \\
	&	5	&	1	&	1	& 2.5881 & 0.9320 & 0.9938 & 15.0061 &	273.5	&		&		\\
	&		&		&&&&&		&	1763.8	&	2205	&	20.00	\\
\end{longtable}
\end{center}
\end{footnotesize}

\section{Predictions of planets according the TB relation}

There are 74 planetary systems with 4 or more planets (see Tables 3 and 4). Among these, 47 planetary systems harbor 4 planets, 18 systems have 5 planets, 7 planetary systems have 6 planets, 1 planetary system contains 7 planets and 1 planetary system contains 8 planets. All these planets have been confirmed. In this sample, 14 systems were detected by the radial velocity technique, 59 systems by transits, and 1 by direct imaging. \\

\noindent This study has shown that the vacancies indicated by the TB relation may be predictions of bodies in planetary systems. Note that they may be planets, asteroids or  groups of small objects. As we mentioned previously, the TBR has an acceptable predictive capacity when using at least 4 planets. So, we applied the TBR to the 74 planetary systems with 4 or more planets, and we found that 25 planetary systems did not present vacancies, 14 systems showed 1 vacancy, 26 planetary systems had 2 vacancies, 7 systems had 3 vacancies and 2 systems showed 4 vacancies (Tables \ref{Table3} \& \ref{Table4}). Tables \ref{Table3} and \ref{Table4} show the values of $p_0$ and $b$ (see equation \ref{exp}) for all the systems, as well as their correlation coefficients R$^2$ and their $\chi^2$. Recently \citet{Kipping2018} applied a different Titius-Bode-type relation to the exoplanetary system Trappist-1  which  gave as a prediction a planet Trappist-1i with a period of 25.345 or 28.699 days, a prediction that is similar to ours (30.08 days). If this prediction is confirmed by future observations the reliability of our method will be increased.\\

\noindent As was shown by the results of our method applied to systems with 5 or more planets, it is possible to suggest that extrapolations of the fits are able to predict new planets, but it is not possible to say how many planets could be found by extrapolating. For this reason, and to be conservative, the numbers of the previous paragraph consider only vacancies but not extrapolations, although some extrapolations could be useful.

\begin{footnotesize}
  
%\end{footnotesize}
\begin{center}
\begin{longtable}{ccccccccccc}
 \caption{TBR fits for systems with 4 planets, predictions and their Minimum detectable mass.}\label{Table3}\\
\hline
 &&&&&&&&&\footnotesize Predicted&\footnotesize Minimum\\
&\footnotesize Planetary&\footnotesize \#&\footnotesize $p_0$&\footnotesize $b$&\footnotesize R$^2$&$\chi^2$&\footnotesize Vacancies&\footnotesize $n$&\footnotesize period&\footnotesize detectable\\
&\footnotesize system&\small planets&(days) & &&&&&\footnotesize  (days)&\footnotesize mass (M$_{\oplus}$)\\
\hline
\endfirsthead
\multicolumn{11}{c}%
{\tablename\ \thetable\ -- \textit{Continued from previous page}} \\
\hline
&&&&&&&&&\footnotesize Predicted&\footnotesize Minimum\\
&\footnotesize Planetary&\footnotesize \#&\footnotesize $p_0$&\footnotesize $b$&\footnotesize R$^2$&$\chi^2$&\footnotesize Vacancies&\footnotesize $n$&\footnotesize period&\footnotesize detectable\\
&\footnotesize system&\small planets&(days) & &&&&&\footnotesize  (days)&\footnotesize mass (M$_{\oplus}$)\\
\hline
\endhead
\hline \multicolumn{11}{c}{\textit{Continued on next page}} \\
\endfoot
\hline
\endlastfoot
%\begin{table}[!ht]
%\footnotesize 
%\begin{center}
%\caption{}\label{tabl3}
%\begin{tabular}{cccccccccccc}
%\hline
%&&&&&&&&\footnotesize Minimum\\
%\footnotesize Planetary&\footnotesize \#&\footnotesize $p_0$&\footnotesize $b$&\footnotesize R$^2$&%\footnotesize Vacancies&\footnotesize $n$&\footnotesize prediction&\footnotesize detectable\\
%\footnotesize system&\small planets&(days) & &&&&\footnotesize  (days)&\footnotesize mass (M$_{\oplus}$)\\
%\hline
1&GJ 3293	&	4	&	8.2758	&	0.4453	&	0.9990 & 0.0619	&	2	&	2	&	20.2	& 11\\
&&	&		&		&		&		&		&	5	&	76.7	&	10	\\
2&GJ 676A	&	4	&	0.7913	&	1.7608	&	0.9938 &85.67	&	1	&	3	&	155.7	&	1	\\
											
3&GJ 876	&	4	&	0.894	&	0.8411	&	0.9959&1.1224	&	2	&	2	&	4.8	&	1	\\
&	&&		&		&		&		&		&	3	&	11.2	&	2	\\
															
4&HD 141399	&	4	&	42.1125	&	0.799	&	0.9997 &1.6394	&	2	&	3	&	462.9	&	11	\\
&	&&		&		&		&		&		&	5	&	2288	&	10	\\
																	
5&HD 215152	&	4	&	4.2152	&	0.2995	&	0.9958 & 0.0482	&	2	&	4	& 14.0	& 2		\\
	&&&		&		&		&		&		&	5	&	18.8	&	2	\\
																	
6&HR 8799	&	4	&	8999.07	&	0.7327	&	0.9981 &135.31	&	0	&		&		&		\\
																	
7&Kepler-106	&	4	&	3.4298	&	0.6455	&	0.9943 & 0.0851	&	0	&		&		&		\\
																	
8&Kepler-107	&	4	&	1.8332	&	0.5088	&	0.9932 &0.3949	&	0	&		&		&		\\
																	
9&Kepler-1388	&	4	&	3.1631	&	0.6284	&	0.9919 & 0.1011	&	0	&		&		&		\\
																	
10&Kepler-1542	&	4	&	2.5255	&	0.1434	&	0.9981 &0.00123	&	2	&	2	&	3.4	&	0.06	\\
	&&&		&		&		&		&		&	4	&	4.5	&	0.06	\\
																	
11&Kepler-167	&	4	&	1.3797	&	0.9426	&	0.9948 &2.0183	&	3	&	4	&	59.9	&	0.22	\\
&	&&		&		&		&		&		&	5	&	153.7	&	0.3	\\
&	&	&	&		&		&		&		&	6	&	394.4	&	0.6	\\
																	
12&Kepler-172	&	4	&	1.2541	&	0.829	&	0.9993 &0.0152	&	0	&		&		&		\\
																	
13&Kepler-176	&	4	&	2.7132	&	0.743	&	0.9971&0.0684	&	0	&		&		&		\\
																	
14&Kepler-197	&	4	&	3.5877	&	0.4929	&	0.9933 &0.04063	&	0	&		&		&		\\
																	
15&Kepler-208	&	4	&	2.9167	&	0.3361	&	0.9926 &0.0318	&	1	&	2	&	5.712	&	0.08	\\
															
16&Kepler-215	&	4	&	6.4685	&	0.3928	&	0.9992 &0.0118	&	2	&	3	&	21.017	&	0.15	\\
&	&&		&		&		&		&		&	5	&	46.105	&	0.2	\\

17&Kepler-220	&	4	&	2.4995	&	0.5938	&	0.9938 &0.1652	&	1	&	3	&	14.841	&	0.12	\\
																	
18&Kepler-221	&	4	&	1.5561	&	0.6216	&	0.9978 &0.0128	&	0	&		&		&		\\
																	
19&Kepler-223	&	4	&	5.218	&	0.3354	&	0.9950 & 0.0175	&	0	&		&		&		\\
																	
20&Kepler-224	&	4	&	1.7662	&	0.6001	&	0.9965 & 0.0398	&	0	&		&		&		\\
																	
21&Kepler-235	&	4	&	1.3749	&	0.8821	&	0.9995 & 0.0189	&	0	&		&		&		\\
																	
22&Kepler-24	&	4	&	2.8267	&	0.3726	&	0.9945 & 0.0357	&	1	&	2	&	6.0	&	0.08	\\
																	
23&Kepler-245	&	4	&	1.4631	&	0.8112	&	0.9988 &0.0407	&	0	&		&		&		\\
																	
24&Kepler-251	&	4	&	2.6403	&	0.6068	&	0.9999 &0.0077	&	2	&	2	&	8.9	&	0.08	\\
 &&&		&		&		&		&		&	5	&	54.9	&	0.2	\\

25&Kepler-256	&	4	&	0.9126	&	0.6202	&	0.9960 & 0.0127	&	0	&		&		&		\\
	
26&Kepler-26	&	4	&	2.2465	&	0.4253	&	0.9964 & 0.1529	&	3	&	2	&	6.0	&	0.05	\\
&	&&		&		&		&		&		&	3	&	8.1	&	0.06	\\
&	&	&	&		&		&		&		&	6	&	28.8	&	0.15	\\
27&Kepler-265	&	4	&	4.3143	&	0.4595	&	0.99999 &0.0009	&	2	&	2	&	10.8	&	0.1	\\
	&&&		&		&		&		&		&	4	&	27.1	&	0.18	\\
	
28&Kepler-282	&	4	&	5.1134	&	0.531	&	0.9919 &0.0920	&	2	&	3	&	17.9	&	0.14	\\
&	&&		&		&		&		&		&	5	&	33.2	&	0.18	\\
	
29&Kepler-286	&	4	&	0.8475	&	0.6949	&	0.9934 &0.1265	&	1	&	4	&	13.7	&	0.1	\\
	
30&Kepler-299	&	4	&	1.2414	&	0.8495	&	0.9989 &0.0402	&	0	&		&		&		\\
	
31&Kepler-304	&	4	&	1.0682	&	0.3765	&	0.9922 &0.0431	&	2	&	2	&	2.3	&	0.005	\\
&	&&		&		&		&		&		&	5	&	7.0	&	0.08	\\
	
32&Kepler-306	&	4	&	2.9263	&	0.452	&	0.9996 &0.0146	&	2	&	3	&	11.4	&	0.1	\\
&	&&		&		&		&		&		&	5	&	28.0	&	0.14	\\
	
33&Kepler-338	&	4	&	7.1186	&	0.3066	&	0.9977 & 0.0203	&	2	&	3	&	17.9	&	0.15	\\
&	&&		&		&		&		&		&	5	&	33.0	&	0.2	\\
	
34&Kepler-341	&	4	&	2.9064	&	0.5442	&	0.9952 &0.1392	&	1	&	3	&	14.9	&	0.12	\\
	
35&Kepler-342	&	4	&	0.9833	&	0.5382	&	0.9981 &0.1671	&	3	&	2	&	2.9	&	0.05	\\
&	&&		&		&		&		&		&	3	&	4.9	&	0.06	\\
&	&	&	&		&		&		&		&	4	&	8.3	&	0.09	\\
	
36&Kepler-37	&	4	&	9.8921	&	0.2736	&	0.9962	& 0.0474 & 	2	&	2	&	17.1	&	0.2	\\
&	&&		&		&		&		&		&	4	&	29.5	&	0.3	\\
	
37&Kepler-402	&	4	&	3.2991	&	0.2026	&	0.9991 &0.0025	&	2	&	2	&	5.0	&	0.06	\\
&	&&		&		&		&		&		&	4	&	7.4	&	0.11	\\
																		
38&Kepler-48	&	4	&	1.4163	&	1.0908	&	0.9939 & 0.7832	&	2	&	4	&	111.2	&	0.4	\\
&	&&		&		&		&		&		&	5	&	331.0	&	0.6	\\
	
39&Kepler-49	&	4	&	1.5954	&	0.4897	&	0.9986 &0.1263	&	1	&	2	&	4.3	&	0.03	\\
																	
40&Kepler-758	&	4	&	3.0082	&	0.4772	&	0.9963 &0.0204	&	0	&		&		&		\\
																	
41&Kepler-79	&	4	&	9.4382	&	0.3521	&	0.9977 &0.1309	&	2	&	2	&	19.1	&	0.2	\\
&	&&		&		&		&		&		&	4	&	38.6	&	0.2	\\
																
42&Kepler-82	&	4	&	1.1865	&	0.7647	&	0.9975 & 0.1196	&	1	&	3	&	11.8	&	0.1	\\
																
43&Kepler-85	&	4	&	5.8508	&	0.369	&	0.9982 &0.0087	&	0	&		&		&		\\
																	
44&Kepler-89	&	4	&	1.6398	&	0.8787	&	0.9969 &0.0618	&	0	&		&		&		\\
																
45&mu Ara	&	4	&	4.045	&	1.0132	&	0.9935 &57.777	&	3	&	2	&	30.7	&	4	\\
&	&&		&		&		&		&		&	3	&	84.5	&	7	\\
&	&&		&		&		&		&		&	6	&	1766.4	&	10	\\
																	
46&tau Cet&	4	&	5.5740	&	1.1572	&	0.9921 & 10.1780	&	0	&		&		&	\\

47&Wasp-47	&	4	&	0.235	&	1.2954	&	0.9939 & 0.8066	&	2	&	4	&	41.6	&	4	\\
&	&&		&		&		&		&		&	5	&	151.4	&	7	\\
\hline 

\end{longtable}
\end{center}
\end{footnotesize}

\begin{footnotesize}
  
%\end{footnotesize}
\begin{center}
\begin{longtable}{ccccccccccc}
 \caption{TBR fits for systems with 5 observed planets or more, predictions and their Minimum detectable mass.}\label{Table4}\\
\hline
& &&&&&&&&\footnotesize Predicted&\footnotesize Minimum\\
&\footnotesize Planetary&\footnotesize \#&\footnotesize $p_0$&\footnotesize $b$&\footnotesize R$^2$&$\chi^2$&\footnotesize Vacancies&\footnotesize $n$&\footnotesize period&\footnotesize detectable\\
&\footnotesize system&\small planets&(days) & &&&&&\footnotesize  (days)&\footnotesize mass (M$_{\oplus}$)\\
\hline
\endfirsthead
\multicolumn{11}{c}%
{\tablename\ \thetable\ -- \textit{Continued from previous page}} \\
\hline
&&&&&&&&&\footnotesize Predicted&\footnotesize Minimum\\
&\footnotesize Planetary&\footnotesize \#&\footnotesize $p_0$&\footnotesize $b$&\footnotesize R$^2$&$\chi^2$&\footnotesize Vacancies&\footnotesize $n$&\footnotesize period&\footnotesize detectable\\
&\footnotesize system&\small planets&(days) & &&&&&\footnotesize  (days)&\footnotesize mass (M$_{\oplus}$)\\
\hline
\endhead
\hline \multicolumn{11}{c}{\textit{Continued on next page}} \\
\endfoot
\hline
\endlastfoot
%\begin{table}[!ht]
%\footnotesize 
%\begin{center}
%\caption{}\label{tabl3}
%\begin{tabular}{cccccccccccc}
%\hline
%&&&&&&&&\footnotesize Minimum\\
%\footnotesize Planetary&\footnotesize \#&\footnotesize $p_0$&\footnotesize $b$&\footnotesize R$^2$&%\footnotesize Vacancies&\footnotesize $n$&\footnotesize prediction&\footnotesize detectable\\
%\footnotesize system&\small planets&(days) & &&&&\footnotesize  (days)&\footnotesize mass (M$_{\oplus}$)\\
%\hline

1&55 Cnc	&	5	&	0.1594	&	1.4738	&	0.9979 &18.3282	&	2	&	2	&	3.0	&	1.8	\\
&&	&		&		&		&		&		&	6	&	1103.7	&	10	\\
															
2&HD 40307	&	5	&	2.0640	&	0.7721	&	0.9972 & 1.9114	&	1	&	5	&	98.1	&	5	\\
																															
3&HIP 41378	&	5	&	10.6414	&	0.3881	&	0.9927 &3.2536	&	4	&	2	&	23.1&	3	\\
&&	&		&		&		&		&		&	4	&	50.3	&	5	\\
&	&&		&		&		&		&		&	5	&	74.1	&	5	\\
&	&&		&		&		&		&		&	8	&	237.4	&	7	\\
																	
4&K2-138	&	5	&	1.5332	&	0.4223	&	0.9999 & 0.0012	&	0	&		&		&		\\

5&Kepler-102	&	5	&	3.7574	&	0.3394	&	0.9920 & 0.1205	&	2	&	4	&	12.4	&	0.1	\\
&	&&		&		&		&		&		&	6	&	21.4	&	0.15	\\
																
6&Kepler-122	&	5	&	3.4068	&	0.4674	&	0.9937 &0.3073	&	1	& 2	& 8.7	&	0.08	\\
																	
7&Kepler-150	&	5	&	1.5435	&	0.7487	&	0.9981 & 0.5182	&	3	&	5	&	65.2	& 0.2		\\
&&	&		&		&		&		&		&	6	&	137.9	& 0.35		\\
&&	&		&		&		&		&		&	7	&	291.5	& 0.5		\\
															
8&Kepler-154	&	5	&	2.1214	&	0.556	&	0.9946 &0.1952	&	1	&	2	&	6.5	&	0.08	\\
																																
9&Kepler-169	&	5	&	2.1922	&	0.4605	&	0.9967 & 0.00015	&	3	&	5	&	21.9	&	0.15	\\
&	&&		&		&		&		&		&	6	&	34.7	&	0.2	\\
&	&&		&		&		&		&		&	7	&	55.1	&	0.2	\\
																
10&Kepler-186	&	5	&	2.2401	&	0.5809	&	0.9995 & 0.0211	&	2	&	5	&	40.8	&	0.18	\\
&	&&		&		&		&		&		&	6	&	72.9	&	0.2	\\
								
11&Kepler-238	&	5	&	1.0262	&	0.6365	&	0.9954 & 0.2194	&	1	&	2	&	3.7	&	0.06	\\
																	
12&Kepler-292	&	5	&	1.4085	&	0.5348	&	0.9946 & 0.0293	&	0	&		&		&		\\
																	
13&Kepler-296	&	5	&	3.2870	&	0.5912	&	0.9995 & 0.0359	&	0	&		&		&		\\
																
14&Kepler-33	&	5	&	4.2327	&	0.2828	&	0.9980 & 0.0609	&	1	&	2	&	8.9	&	0.09	\\
																
15&Kepler-444	&	5	&	2.8045	&	0.2523	&	0.9969 & 0.0063	&	0	&		&		&		\\
																
16&Kepler-55	&	5	&	1.5356	&	0.4786	&	0.9939 & 0.1071	&	2	&	3	&	6.4	&	0.07	\\
&	&&		&		&		&		&		&	5	&	16.8	&	0.11	\\
																
17&Kepler-62	&	5	&	3.1299	&	0.6239	&	0.9959 & 1.4485	&	2	&	4	&	38.0	&	0.2	\\
&	&&		&		&		&		&		&	5	&	70.9	&	0.2	\\
																	
18&Kepler-84	&	5	&	2.7741	&	0.3890	&	0.9981 & 0.1898	&	2	&	2	&	6.0	&	0.07	\\
&	&&		&		&		&		&		&	5	&	19.4	&	2	\\

19&GJ667C	&	6	&	4.5715	&	0.4391	&	0.9975 & 1.09185	&	3	&	2	&	11.0	&	2	\\
&	&&		&		&		&		&		&	3	&	17.1	&	2	\\
&	&&		&		&		&		&		&	8	&	153.3	&	4	\\
																	
20&HD 10180	&	6	&	1.51	&	1.1855	&	0.9927 & 45.073	&	0	&		&		&		\\
																
21&HD 219134	&	6	&	1.1418	&	0.9338	&	0.9953 & 36.5276	&	2	&	6	&	309.6	&		\\
&	&&		&		&		&		&		&	7	&	787.8	&		\\

22&HD 34445	&	6	&	26.2016	&	0.5344	&	0.9982 & 13.6713	&	4	&	2	& 76.3		&	5	\\
&	&&		&		&		&		&		&	5	&	379.1	&	10	\\
&	&&		&		&		&		&		&	8	&	1883.8	&	15	\\
&	&&		&		&		&		&		&	9	&	3214.5	&	18	\\

23&Kepler-11	&	6	&	6.2963	&	0.412	&	0.9939 & 0.6974	&	1	&	6	&	74.6	&	0.3	\\
																
24&Kepler-20	&	6	&	1.8483	&	0.6014	&	0.9949 & 0.7062	&	0	&		&		&		\\
																
25&Kepler-80	&	6	&	0.6739	&	0.3840	&	0.9991 & 0.0309	&	2	&	2	&	1.5	&	0.001	\\
&	&&		&		&		&		&		&	3	&	2.1	&	0.005	\\
																
26&TRAPPIST-1	&	7	&	1.0486	&	0.4263	&	0.994	&	0.2214&0	&		&		&		\\

27&Kepler-90	&	8	&	4.0530	&	0.4387	&	0.9975 & 1.4610	&	2	&	4	&	23.4	&	0.14	\\
&	&&		&		&		&		&		&	5	&	36.3	&	0.2	\\
\hline 

\end{longtable}
\end{center}
\end{footnotesize}

\section{Statistical Analysis}

\begin{figure}[ht!]
  \begin{center}
    \includegraphics[scale=0.65]{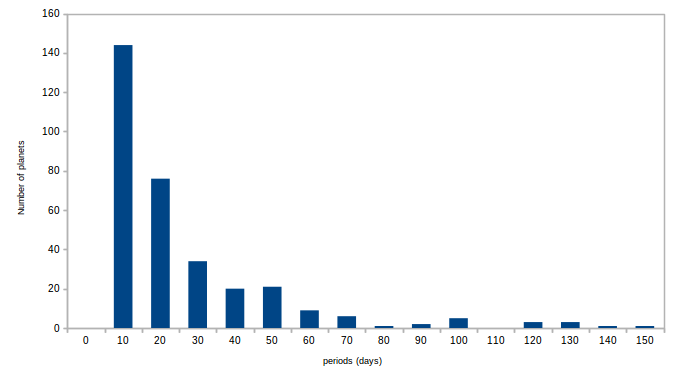}
    \caption{Histogram of observed periods of all known exoplanets. The data were taken from the catalog   http://exoplanet.eu at the time of writing. The periods plotted are those of all confirmed by the different detection methods.}
    \label{periodsobserved}
  \end{center}
\end{figure}

\noindent In order to assess the statistical significance of our results, we performed a series of fits on randomly generated artificial systems. We generated 900 random systems, namely 300 systems with five, six and seven planets, respectively. Each of the sets of 300 systems for five, six or seven planets was generated with a uniform distribution of periods for 100 systems, an exponentially increasing distribution for the second 100 systems and an exponentially decreasing distribution for the last 100 systems.   We considered only periods between 1 and 100 days to reflect the observed period distribution of the great majority of exoplanets (Figure \ref{periodsobserved}). \\

\noindent We applied our method to each one of the 900 random systems in exactly the same way as we did for the real systems. The results are shown in Table \ref{tab:1.1}. We found that in all of the artificially generated sets there were fewer successful cases than unsuccessful ones, which is contrary to what we found for the real systems.  In addition, the larger the number of planets the fewer successful cases were found. This is just the opposite to what we found applying our method to the real systems. In particular, we expected that random systems with an exponentially increasing distribution of periods (a TB-like distribution, similar to that of real systems) would favor successful cases; in fact these showed  results worse than those for systems with an  exponentially decreasing period distribution. \\

\begin{table}[hb]
  \caption{Statistical Results}
  \label{tab:1.1}
  \begin{center}
    \begin{tabular}{ccccccc}
    \hline
      Number & Result &Uniform  & Exponentially  & Exponentially  &Real Sample \\
      of Planets & &Distribution & Increasing & Decreasing  & of \\
      &&&Distribution&Distribution& Exoplanets\\
       \hline
   5& Successful & 33 \% & 26 \% &42\% &72\% (13/18) \\
   &Unsuccessful &67\%& 74\%& 58\%& 28\% (5/18)\\
    &            &    &     &     &            \\
   6& Successful & 5 \% & 17 \% &27\% &85\% (6/7) \\
   &Unsuccessful & 95\%& 83\%& 73\%& 15\% (1/7) \\
     &            &    &     &     &            \\
   7& Successful & 1 \% & 2 \% &28\% &100\% (1/1) \\
   &Unsuccessful & 99\%& 98\%& 72\%&  \\
   \hline
       \end{tabular}
  \end{center}
\end{table}

\noindent From these tests we can conclude that the real exoplanetary systems that satisfy the TB relation are systems whose periods do not stem from a random distribution. Moreover, these results indicate that the configurations of real planetary systems reflect the fact that the bodies in them are not independent, but rather they are influenced by the presence of the other bodies in the system.  Thus, their relative positions are either a consequence of the formation mechanism or due to mutual interactions. Indeed, we obtain a precision of only 26\% when we draw the planets (independently) from an exponentially increasing distribution, as compared with a precision of 72\% for the real systems. Hence, the predictions of our method should be reliable. 

\section{Discussion}

\noindent Among the 27 systems with 5 or more planets (Tables \ref{Table1} and \ref{Table2} ), we found our method to be  successful for 21 systems. We found a total of 32 conserved vacancies. Hence, we posit that these vacancies predict the periods of as yet undiscovered objects. Among the remaining 6 systems studied, the vacancies were not conserved. However, we must take into account that the fits depend on the determination of the orbital elements and the errors associated with the observations. An example that shows that the fits depend on the quality of the data can be found in \citet{Poveda2008} and \citet{Lara2012}. In their 2008 work, an orbital period of 2.817 days for 55 Cancri-e was used, and a vacancy at $n = 5$ with an orbital period of 1130 days was predicted. \citet{Dawson2010} reported a correction to the orbital period of 55 Cancri-e. With this new datum, \citet{Lara2012} predicted two vacancies at $n = 2$ (3.038 days) and $n = 6$ (1103.64 days). This example shows that vacancies may change as better periods are obtained. In addition, we note  that as the determination of periods improves, our method becomes more successful.  \\

\noindent As we just mentioned, our method was successful for 21 exoplanetary systems. The average error between the observed planets and the predictions was $9.84\%$. The best case was Kepler-296, with an error of $0.63 \%$. Among the successful cases, the worst was Trappist-1, with an error of $29.49\%$ when four planets were used. However,  when more planets were considered the errors decreased to $1.93\%$. In general, the errors decreased when using more planets in the fit, with the exception of GJ 667C, HD 34445, Kepler-11 and Kepler-80, where the predicted error in the period of a planet increased by $0.97\%$ on average.The least successful example was Kepler 150, for which 9 vacancies were predicted, but only 3 were conserved.\\

\noindent To assess the goodness of the fits, we calculated the $\chi^2$ for all the possible combinations between planets and vacancies (8th column in Tables \ref{Table1} and \ref{Table2}). We noticed that when our criterion ($R ^ 2 = 0.9920$) was fulfilled, the $\chi ^ 2$ was a minimum. \\ 

\noindent According to \citet{Bovaird2013}, the more compact exoplanetary systems are, the better they are described by a TB relation. Following this idea, we assessed whether or not the success of a system depended on its compactness. To do this, we calculated the difference of the periods between contiguous planets (Figure \ref{fig3} ). This is a measure of how far planets are from each other. Figure \ref{fig3}, shows that there is no evident difference between the successful cases (Figure \ref{fig3} a) and the unsuccessful cases (Figure \ref{fig3} b), that is, the success of the TB relation is not affected by the compactness of the system.\\

\noindent According to the results of the previous section, the more compact systems simulated by random systems with an exponentially decreasing distribution of periods show a larger number of  successes than the other cases. This may be due to the fact that the periods (and thus the physical spacing of the planets) are so small that they approximately follow a linear relation, which makes it easier to obtain good fits. For non-compact, (i.e., physically larger systems with larger periods), the fits are worse than for compact systems. \citet{Bovaird2013} found a similar result. 
In conclusion, Table 5 shows that the distribution of periods of the real systems is not random, and thus it does not resemble that of the random systems with which we compared it.     \\

\begin{footnotesize}
  
\begin{figure}[h!]
  \begin{center}
    \includegraphics[scale=0.62]{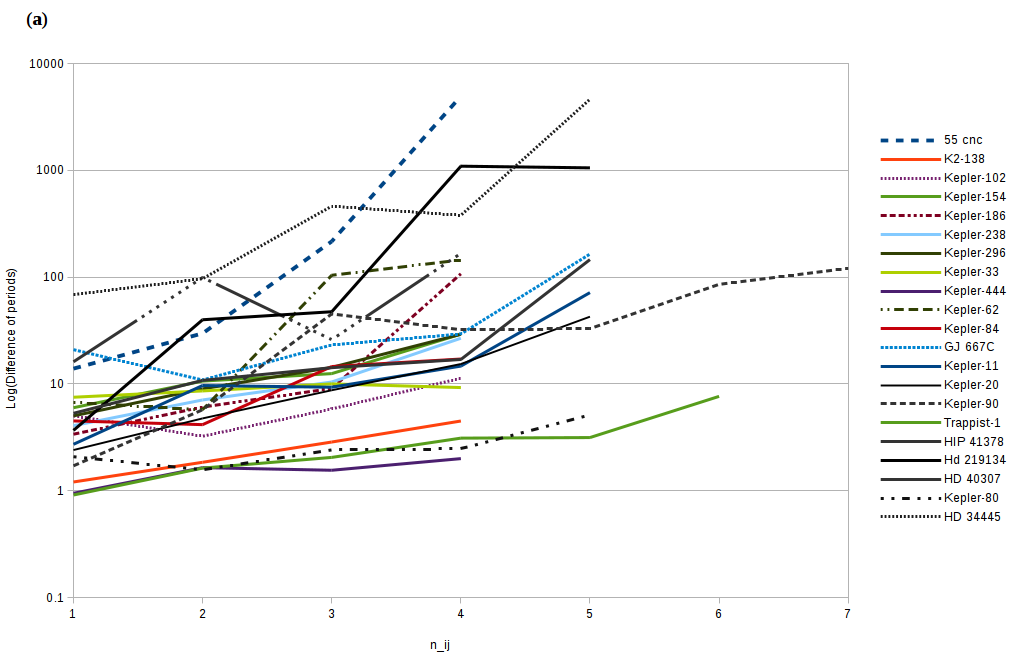}\\
    \includegraphics[scale=0.62]{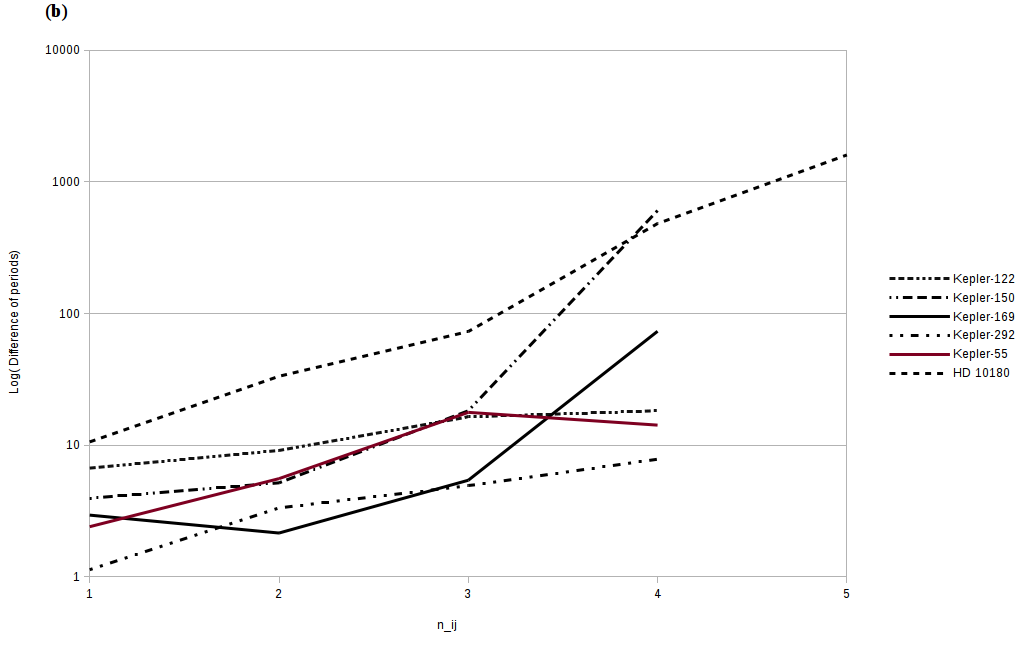}
        
    \caption{Distribution of the differences of periods between contiguous planets. (a) Successful cases, and (b) Unsuccessful cases.}
    \label{fig3}
  \end{center}
\end{figure}

\end{footnotesize}

\noindent Although our sample still has a small number of systems, we can conclude that the TB relation, together with our criterion, is able to predict new planets with a precision of $78\%$. This percentage could increase when more and better data become available. For now, it is likely that the 32 vacancies found in the 21 successful systems correspond to real bodies (planets, asteroids or belts). In fact,  from the moment that we started to analyze these systems (Poveda \& Lara 2008) to the present, oth the increase of available and better data have improved the precision of our method; indeed, it has increased from $56\%$ to $78\%$.\\

\noindent Applying our method to the 74 systems with 4 planets or more, we found 95 vacancies  (Tables \ref{Table3} and \ref{Table4}). These may be bodies that have not yet been discovered because of detection limits. We can estimate the mass of the predicted objects by means of the minimum planetary mass detectable (depending on the detection technique and program). Figure 9.14 of Rauer and Erikson (\citet{Dvorak2008}) shows the limits of detection of masses and distances of planets from the parent star depending on the observation method. For the exoplanetary systems detected by transits, we found that the vacancies could indicate planets with masses similar to Mercury and Ceres,  but they could also be a set of objects similar to our main asteroid belt. In general, the radial velocity method finds planets more massive than the transit method. So, in the systems detected by radial velocity, the objects predicted by the vacancies should have masses smaller than 5 terrestrial masses.\\

\noindent Using our predicted data, we can analyze how feasible it is that these bodies could be stable in their corresponding locations in exoplanetary systems. For this purpose we can use the dynamic spacing criterion $ \Delta $ which is defined based on the radius of the planets (equation \ref{ecdelta}). This criterion was calculated by \citet{Gladman1993} and \citet{Chambers1996}. They mention that, as a result of orbital simulations of multiple systems, two contiguous planets become unstable when $ \Delta <10 $. Using this criterion, we calculated $ \Delta $ for all systems with known planetary masses, a total of 17 systems. We used the criterion both for the observed planets and for the observed planets plus the vacancies predicted by the TB relation. For the vacancies, we used a threshold mass according to the detection method for each system.\\

\begin{equation} \label{ecdelta}
\centering
\Delta= \frac{2(M_*)^{1/3}(p_{2}^{2/3}-p_{1}^{2/3})}{(p_{2}^{2/3}+p_{1}^{2/3})(m_1+m_2)^{1/3}}
\end{equation}

\begin{figure}[ht]
  \begin{center}
    \includegraphics[scale=0.7]{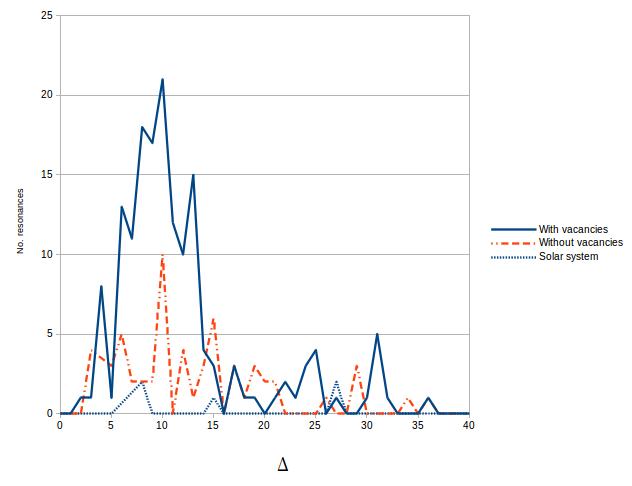}
    \caption{Total number of resonances}
    \label{figdelta}
  \end{center}
\end{figure}

\noindent We found that when $ \Delta $ is calculated using only the observed planets, the number of pairs of planets with $\Delta > 10$ is greater than that found when the vacancies predicted by TB were included. However, we noted that when we applied this criterion to the Solar System we found for Ceres and Jupiter a $\Delta$ less than 10, suggesting that they are unstable.  However Jupiter and Ceres are actually stable because they are in resonance with other bodies. The same is likely to be true for some exoplanetary systems, because we found that when vacancies are considered the number of planets and vacancies with $\Delta$ $< 10$ increases as compared to the case with no vacancies (see Figure \ref{figdelta}). However, the number of resonances between planets does increase when vacancies are considered. Hence, the existence of objects in the vacancies would imply stability due to resonances.\\

\noindent To finish this section, we want to emphasize that the aim of this paper has been to assess the capability of the TB relation to predict bodies not yet discovered in  exoplanetary systems. To test the robustness  of our results, we applied our method to 900 randomly generated artificial systems, as  described in Section 5. According to Table 5, the artificial systems showed fewer successes  than the real systems.  In addition, the systems ones with a larger number of exoplanets showed fewer successes, which is contrary to what we found for the real systems. This is a further indication that the distribution of periods of the real systems does not stem from a random distribution.

\section{Habitable Zone}

\noindent In order to verify whether some of the bodies predicted by our method can be habitable planets  we used the conservative definition of the Habitable Zone of Kasting et al. 1993. With the luminosity of the primary star and its spectral type, we calculated the limits of the Habitable Zone for all 74 exoplanetary systems with 4 o more planets.\\

\noindent Among the 21 successful cases we found, there are 7 planets within the Habitable Zone: 55 Cnc f, Kepler-296 f, Kepler-62 e, GJ 667C c, GJ 667C f and Trappist-1 e and f. All of them correspond to observed planets. \\
 
\noindent After applying our method to the full sample of 74 exoplanetary systems, we found 4 vacancies within the Habitable Zone of the respective primary star. The predictions are: Kepler-167 with a period of 153.66 days, Kepler-48 with a period of 330.98 days, Kepler-150 with a period of 291.475 and Kepler-186 with a period of 72.89 days. However, we consider  Kepler-150 to be  an unsuccessful case, so this prediction is less certain. Kepler-167 and Kepler-48 are systems with 4 observed planets, so our predictions have a precision of $78 \%$. We note that Kepler-186 is one of our successful systems, so we think that our prediction has a high probability of being confirmed.

\section{Conclusions}

\noindent The TB relation has been controversial due to the lack of a physical explanation for it. However, considering that the TB relation can be represented by equation 2, it turns out that all the known exoplanetary systems follow such a relation. The usefulness of such a relation to predict new planets has not been quantitatively assessed until the present work.  \\

\noindent We found that at least 4 observed planets are needed to obtain fits with an acceptable correlation coefficient and minimum errors. We also found that the more planets we use to fit the data, the greater will be the predictive capacity of the TB relation according to the proposed method. We found that the TB relation is able to predict new planetary bodies in a system with a precision of 78\%. In other words, in 78 \% of the cases the TB relation successfully predicted new planets, whereas in 22 \% of the cases it failed. This can be compared with the rates we find when the planets are randomly (but independently) drawn from an exponentially increasing population, which are 26 \% of successful cases and 74 \% of failed cases.  We interpret this as a consequence of the fact that true planets interact with each other. It is likely that with new and better data the TB success rate will increase.\\

\noindent In addition, we showed that real systems behave differently from randomly generated systems; for the comparison we used three different period distributions for the artificial systems. This means that the distribution of periods of the real systems does not stem from a random distribution, and it provides further evidence that the planets interact with each other.\\

\noindent According to our results, it is probable that Kepler-186 has a new planet or an asteroid belt within its habitable zone.

%%%%%%%%%%%%%%%%%%%%%%%%%%%%%%%%%%%%%%%

\begin{ack}
We thank the referee for a constructive report and very interesting suggestions, which improved the paper.  Our interest in the TB relation was originally inspired by A. Poveda, to whom we are grateful. PL gratefully acknowledges support from Grant 378670 of Consejo Nacional de Ciencia y Tecnologia (CONACyT) 

\end{ack}

\appendix 
\section{Tables} \label{Appendix}

%\selectlanguage{USenglish}

%%%
% See the manual for the detail.
%%%
%\begin{thebibliography}{}
% Journals(e.g. A\&A,ApJ,AJ,NMRAS,PASP ...)
% Authors, Year, Journal, Vol#, Page#
% Journal Title Abbreviation >> http://www.asj.or.jp/pasj/Jabb.html

\end{document}